\documentstyle[fullpage,epsf,twocolumn]{article}
\def\<{\langle}
\def\>{\rangle}
\twocolumn
\title{Paramagnetic structure for the soliton of the $30^\circ$ partial dislocation in silicon}
\author{G\'abor Cs\'anyi, Sohrab Ismail-Beigi and T.A. Arias\\
Department of Physics\\
Massachusetts Institute of Technology\\
Cambridge, MA 02139
}

\date{\ }

\begin{document}

\maketitle

\begin{abstract}
Based on {\em ab initio} calculation, we propose a new
structure for the fundamental excitation of the reconstructed
30$^\circ$ partial dislocation in silicon.  This soliton has a rare
structure involving a five-fold coordinated atom near the dislocation
core.  The unique electronic structure of this defect is consistent
with the electron spin resonance signature of the hitherto
enigmatic thermally stable R center of plastically deformed
silicon. This identification suggests the possibility of an
experimental determination of the density of solitons, a key defect in
understanding the plastic flow of the material.  \break PACS {\bf 71.15.Mb}
\end{abstract}

Dislocations are central to the understanding of the mechanical
response of materials.  The mechanical behavior of any crystalline
material is determined by a hierarchy of crystalline defects of
successively lower dimension.  Grain boundaries are two dimensional
defects that control the evolution of the microstructure of the
material.  The creation and motion of dislocations, which are one
dimensional extended topological defects of the lattice, mediate the
plastic response of a crystal to external stress.  In silicon, which
has a bipartite lattice, the primary mobile dislocations are the screw
and the $60^\circ$ dislocation, which dissociate into more primitive
one-dimensional partial dislocations bounding stacking faults.  The
mobility of dislocations in high Peierls barrier materials such as
silicon is effected by the motion of complex zero-dimensional local
defects known as {\em kinks}, where the dislocation center skips from
one row of atoms to a neighbouring row.  The low energy kinks along
the $30^\circ$ partial dislocation in silicon have been shown to also
involve a composite structure where kinks bind with zero-dimensional
soliton excitations in the reconstructed ground state of the
dislocation core.  These soliton excitations are also known as
``anti-phase defects'' (APDs)~\cite{Bulatov}.

Here we report the results of an {\em ab initio} study exploring the
lattice and electronic structures, excitation energy, and the density
of these APDs which are the simplest, lowest energy, fundamental
excitations of the dislocations in the hierarchy ultimately leading to
the macroscopic mechanical behavior of the solid.  The soliton is
associated with an atom in the dislocation core which is not part of a
reconstructed dimer.  In the simple, conventional picture, this atom
(henceforth to be referred to as the ``soliton atom'') only has three
bonds and therefore an unpaired electron.  This simple model, however,
does not lead to predictions consistent with any of the observed
electron spin resonance (ESR) signals associated with plastically
deformed silicon.

We propose a new structure of the soliton.  We find that the ground
state of the soliton has an unexpected structure with electronic
states which {\em are} consistent with the most stable ESR center in
plastically deformed silicon, the only one which remains after careful
annealing.  The reason why the natural connection between this center
and the lowest energy excitation of the dislocation core has not been
made previously is that the observed ESR center has a highly unusual
symmetry.  In support of our theory for the structure of the soliton,
we gather here several pieces of evidence from both reports of ESR
results and our own {\em ab initio} calculations.  The final combined
{\em ab initio}--experimental identification which we make allows for
the possibility that future more precise measurements of the ESR center
density could be used to determine experimentally the soliton density,
a key physical quantity in the process of deformation.

\begin{figure}
\epsfxsize = 2.5in
\centerline{\hfil\epsffile{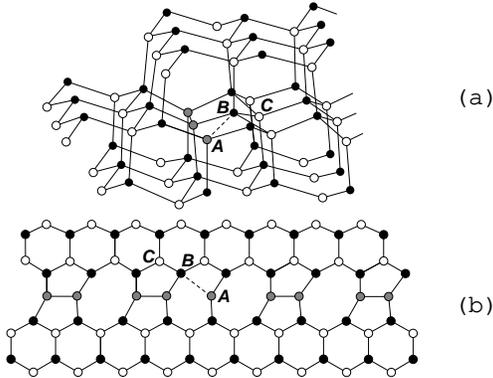}\hfil}
\caption{\vbox to 7mm{} (a) Three dimensional view of a
$30^\circ$ partial dislocation, with a soliton domain wall, (b)
schematic drawing of the same in the (111) plane. The grey circles
represent the atoms in the central core of the dislocation.}
\label{fig:disloc}
\end{figure}

Figure \ref{fig:disloc} reviews the basic geometry of the $30^\circ$
partial dislocation studied in this work.  The dislocation is the
one-dimensional boundary defining the edge of a half-planar (111)
stacking fault.  Atoms in the central core of the dislocation (shaded
gray in the figure) are connected to the bulk with only three bonds
per atom.  The dislocation undergoes a reconstruction whereby the core
atoms pair up in dimers forming intra-core bonds and thus become
four-fold coordinated.  This reconstruction breaks the original
translational symmetry along the core and doubles the primitive repeat
distance along the core axis.  Associated with this broken
translational symmetry is a low energy soliton defect where, by
creating a single unpaired core atom, the system may change along the
line from one of the two symmetry related degenerate ground state
phases to the other.  Because the soliton atom is expected to have a
dangling bond, it is natural to look for an ESR signal for this
defect.  The relatively low energy we expect for such an excitation
leads us to expect a relatively large equilibrium population at
silicon annealing temperatures ($\sim 900$K) and therefore that the
ESR signal would not anneal out as quickly as other signals associated
with the formation and motion of dislocations.  We further would
expect this signal to be detected in all systems which contain
$30^\circ$ partial dislocations.

Indeed, it was discovered over thirty years ago that plastically
deformed silicon gives a wide variety of ESR signals \cite{ESR}.  (For
an excellent review, see \cite{Alexander}.)  Out of the dozens of ESR
centers, four have been identified as associated with the $30^\circ$
dislocation core.  The literature refers to these centers as Si-K1,
Si-K2, Si-Y and Si-R.  The K1 and K2 defects have been identified to
be electronic excitations of the same structural defect.
Kisielowski--Kemmerich \cite{KK} made the currently accepted
identification of the K and Y defects.  It is well known that the
first three of the aforementioned ESR signals anneal out over the
(temperature dependent) time-scale of about an hour \cite{Alexander}.
Only one signal remains, the one labelled R \cite{Y-R,Omling}.  This
center is ``thermally stable'' (it does not anneal out) and is
observed even at high deformation temperatures ($> 900^\circ$ K) where
the other signals anneal out too quickly to be observed.  The R signal
is the residue of Y after annealing, and is very similar to Y in its
diminished anisotropy and large width \cite{Y-R}.

Many authors have studied excitations of dislocations in silicon
\cite{Bulatov,Nunes,Bennetto,Spence}.  The nature of the ESR centers
and the soliton excitation energy of the 30$^\circ$ partial
dislocation, however, have yet to be addressed with modern {\em ab
initio} techniques.  In order to investigate the electronic structure
of low energy excitations of the $30^\circ$ partial dislocation core,
we embarked upon a density functional study of the system.

To prepare approximately relaxed initial ionic configurations with the
correct bonding topology, we first relaxed lattices containing
dislocation cores using the Stillinger--Weber (SW) inter-atomic
potential\cite{SW}.  While doing this, we discovered that the soliton
atom moves out of line with respect to the dislocation core.
To probe this interesting possibility further, we carried out
calculations within the plane wave total energy density functional
approach\cite{RMP}.  To describe the electron-electron interactions we
used the Perdew-Zunger\cite{PerdewZunger} parameterization of the
Ceperly-Alder\cite{CeperlyAlder} exchange-correlation energy of the
uniform electron gas.  To describe the electron-ion interactions we
used a non-local pseudopotential of the Kleinmann--Bylander
form\cite{KB}.  The electronic wave functions were expanded in a
plane-wave basis up to a cutoff of $8$ Ry.

All super-cells used in this study have the same size in the plane
perpendicular to the (110) dislocation axis.  Two partial dislocations
of equal but opposite Burgers vectors at a separation of $14$ \AA\ cut
through this plane.  Following Bigger {\em et al.} \cite{Bigger}, the
lattice vectors are arranged so that the periodic dislocation array
has a quadrupolar arrangement.  Each cell contains forty-eight atoms
per bilayer stacked along the dislocation core direction.  To
calculate the excitation energy of the soliton, it is also necessary
to calculate the energy of the perfectly reconstructed dislocation.
However, the smallest super-cell which is commensurate with both
structures contains six bilayers ($288$ atoms).  It is possible,
however, to reduce the computational time by using two different
super-cells.  For the reconstructed case, the super-cell contains two
bilayers along the dislocation line, while the soliton structure
contains three, ($96$ and $144$ atoms, respectively).  The lattice
vectors were obtained by relaxing a completely reconstructed
dislocation within the SW model in the $96$ atom cell.  The three
bilayer cell was then obtained from this by multiplying the lattice
vector that points along the dislocation axis by ${\scriptstyle
3\over2}$.

\begin{figure}
\epsfxsize = 2.5in
\centerline{\hfil\epsffile{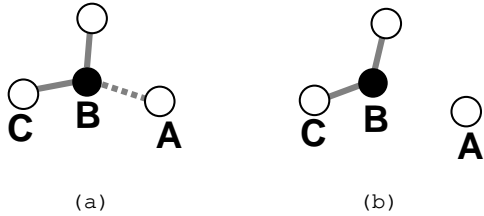}\hfil}
\caption{\vbox to 7mm{} Coplanar atoms near the soliton: (a) in the proposed ground
state for the soliton, (b) conventional structure with the soliton
atom in line with the remaining atoms in the dislocation core.  The
notation for the atoms is the same as in Fig. \ref{fig:disloc}: atom A
is the soliton atom, B is the five-fold coordinated bulk atom (see
text), C is a bulk atom bonded to B opposite from A.}
\label{fig:shift}
\end{figure}

To minimize errors from k-point sampling, basis set truncation and
super-cell effects, we compute differences of energy differences, as
follows.  For each super-cell, we generated a completely
unreconstructed configuration where all the core atoms have only three
bonds.  These structures can be realized in both super-cells, and thus
serve as the reference point. The final excitation energy is the
difference between the deviations in the energy from the
unreconstructed structure in each cell.  By keeping the lattice
vectors fixed throughout the calculations, we more closely simulate
the environment which widely separated solitons would experience along
a reconstructed dislocation line.  To test for the impact of strain
effects, we repeated all calculations using the corresponding lattice
vectors of a bulk silicon system at the {\em ab initio} lattice
constant. Our results did not change significantly.

To ensure maximum transferability of results between the two cells, the
calculations employed k-point sets which give identical sampling of
the Brillouin zone for the two super-cells: $\{(0, 0, \pm 1/4)\}$ for
the three bilayer cell and $\{(0, 0, \pm 1/6), (0, 0, 1/2)\}$ for the
two bilayer cell. To find relaxed structures, we moved all ions
along the Hellmann--Feynman forces until the ionic forces were less
than $0.02$ eV/\AA.  Typically, this was accomplished in $40$ ionic
steps, where between ionic steps we made $10$--$15$ electronic
relaxation steps using the analytically continued functional
approach\cite{ACprl}.

Figure \ref{fig:shift} shows the projection in the $(110)$ plane of
our {\em ab initio} results for the structure of the soliton.  (The
same atoms are shown in Figure \ref{fig:disloc} in full structural
context.)  The structure on the left is our prediction for the ground
state.  In this configuration, there is a five-fold coordinated bulk
atom (B) in the immediate neighbouring row to the dislocation core.
Its new, fifth neighbor is the soliton atom (A).  Our {\em ab initio}
results show that the conventional structure on the right (generated
by keeping the soliton atom collinear with the dislocation core) is
not only higher in energy but also spontaneously decays into the
ground state on the left.

The {\em ab initio} excitation energy of the soliton is
$0.65~\pm~(\approx0.2$)~eV, where we attribute most of the uncertainty
to the uncontrolled local density approximation and super-cell
effects.  This energy corresponds to a density $\rho =
{\scriptstyle{1\over2}}e^{-E/kT}$ in the range of $10^{-5}$ to
$2\times10^{-3}$ solitons per core atom at $900$ K, which is
consistent with the observed densities of the ESR centers: the R
center represents a ``fraction'' of the density of the Y center, which
is estimated in the experimental literature to be about $0.01$ per
core atom (Table II in \cite{Omling}, Table 2 in \cite{Ydens}, and
\cite{Alexander}).  (The factor of one half in $\rho$ comes from the
fact that in a given phase of the ground state, only half of the core
atoms represent possible soliton sites.)  Given the exponential
sensitivity of the density, we find this agreement encouraging,
particularly as the energy of soliton is very low compared typical
point defects in silicon.

\begin{figure}
\epsfxsize = 3.5in
\centerline{\hfil\epsffile{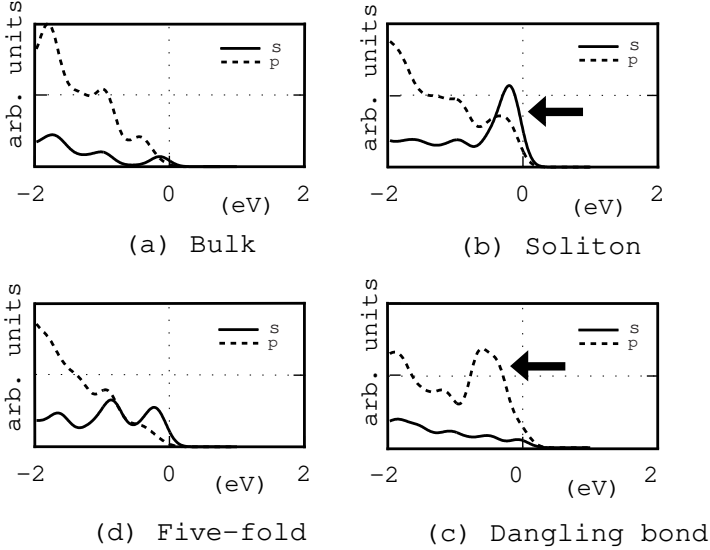}\hfil}
\caption{\vbox to 7mm{} Local density of states, calculated by acting on the filled
bands with the Kleinmann--Bylander projectors centered at (a) an atom
deep in the bulk , (b) the soliton atom in its proposed ground state,
(c) a conventional soliton atom with a dangling bond, (d) the quasi
five--fold coordinated atom.  The horizontal axis is the energy (eV),
the scale of the vertical axis is arbitrary but the same for all four
plots.  Solid and dashed lines represent densities in the {\em s} and
{\em p} channels respectively.}
\label{fig:ldos}
\end{figure}

A great advantage of {\em ab initio} calculations, beyond their
reliability, is that they also yield the electronic states, in
particular giving information about their spatial symmetry.  In Figure
\ref{fig:ldos} we plot the angular momentum decomposition of the local
density of states as obtained from the Kleinmann--Bylander projections
of the electronic eigenstates.  Panel (a) shows, for an atom far from
the core, the familiar concentration of p--like states at the top of
the valence band.  To explore the nature of the soliton state, we
compare this to the local densities of states for the soliton atom in
the proposed (\ref{fig:ldos}b) and conventional (\ref{fig:ldos}c)
configurations.  We also plot the local density of states for the
quasi-fivefold coordinated atom (\ref{fig:ldos}d).  The appearance of
the peak near the top of the valence band in the {\em s} channel of
the soliton atom (and the corresponding diminution in the {\em p}
channel) in its ground state (\ref{fig:ldos}b) shows that the state
associated with the soliton is much less anisotropic than the simple
dangling p--like bond on the soliton atom in the conventional picture
(\ref{fig:ldos}c).  We further note an enhancement at the same energy
in the {\em s} channel of the quasi-fivefold coordinated atom
(\ref{fig:ldos}d), which indicates that the unpaired electron is
shared between this atom and the soliton atom.  Defects in the
dislocation core therefore need not be associated with strongly
directional electronic states, as has been previously assumed in
identifications of ESR centers.

There is a direct connection between the symmetry of the electronic
state of an unpaired spin and the symmetry of the corresponding ESR
signal as described by its effective $g$ tensor.  The off diagonal
elements of this tensor involve a sum over matrix elements of the form
$\<\phi_0|{\bf L}_i|\phi_e\>\<\phi_e|{\bf L}_j|\phi_0\>$, where
$\phi_0$ is the unpaired state and $\phi_e$ are the excited
states\cite{gtensor}.  In general $\phi_0$ can be broken into angular
momentum components (as in Figure \ref{fig:ldos}), of which the $s$
wave component makes no contribution to the preceding matrix elements.
In general, then, we expect the anisotropy of $g$ to be proportional
to the population of the $p$ channel.  (Higher angular momentum
components are negligible for filled states in silicon.)  This
population, the area under the peak associated with the unpaired
electron in the {\em p} channel, drops by about a factor of two as the
soliton moves from its symmetrical dangling bond configuration
(\ref{fig:ldos}c) to our proposed state (\ref{fig:ldos}b).  The
literature contains qualitative observations of the decreased
anisotropy of the R signal and one quantitative comparison which comes
from measurements of the Y signal, of which the R is the residual
after annealing.  In \cite{KK}, this anisotropy is compared directly
with that of the K1,2 centers, which have typical dangling bonds, and
is shown to be less by about a factor of two, in agreement with our
electronic structure results.

\begin{figure}
\epsfxsize = 2.5in
\centerline{\hfil\epsffile{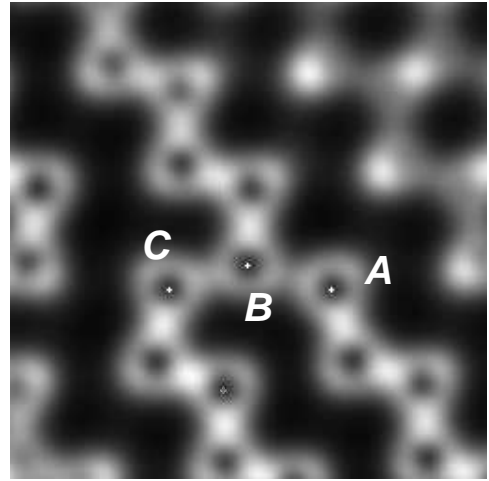}\hfil}
\caption{\vbox to 7mm{} Two dimensional slice of the total charge density from our
{\em ab initio} calculation, through the plane containing the atoms A,
B, C, as labelled in the previous figures. The new A--B bond is nearly
as pronounced as the bulk B--C bond, which is clearly weakened
compared to the other (vertical) bulk bond of atom B seen in the
figure.}
\label{fig:dens}
\end{figure}

To explore the nature of the bonding near the soliton atom, we plot
the total valence charge density in Figure \ref{fig:dens}, which shows
that the soliton atom (A) makes a weak bond with the neighbouring
five--fold coordinated atom in the bulk (B).  The new bond (A--B) of
the bulk atom is very similar to a now weakened but previously
existing bond in the spatially opposite direction (B--C). Such
five-fold coordinated structures have been considered previously in
silicon by Pantelides \cite{Pant} and more recently by Duesbery et
al.\cite{Duesbery}.  Through {\em ab initio} studies it was
demonstrated that such five-fold defects in amorphous silicon should
show similar anomalies in the angular momentum decomposition of the
local density of states\cite{floatDFT}.

In conclusion, we have presented {\em ab initio} results indicating
that the ground state of the soliton has an unusual structure
involving a five-fold coordinated atom and a correspondingly unusual
electronic structure.  The excitation energy we calculate for this new
defect corresponds to a thermal equilibrium density which is
compatible with the observed density of the R center, which is the
only thermally stable paramagnetic center associated with the
$30^\circ$ partial dislocation.  In line with our notion of the
soliton being the fundamental excitation of the reconstructed
dislocation core, the R center is observed independent of the method
of deformation and in proportion to the dislocation density.  Our
calculations show that the electron state of the soliton has a reduced
anisotropy compared to that of a simple dangling bond which
corresponds in magnitude to the puzzling reduction in anisotropy of
the ESR signal of the R center.  In the ground state structure which
we propose, the soliton atom makes a weak bond with a neighbouring
bulk atom and thus gives rise to an amorphous-like bonding
arrangement.  This could explain in part the oft noted similarity of
the ESR signature of the R center to that of amorphous silicon.  Based
on the above arguments and results, we propose that the domain walls
in the reconstruction of the $30^\circ$ partial dislocation and the R
centers observed in ESR experiments are one and the same.  Any viable
competing theory which does not identify the R signal with the soliton
must both predict a more plausible intrinsic excitation corresponding
to the the R center and explain why the unpaired electron of the low
energy soliton is not detected in ESR experiments.

\begin{center}{\bf Acknowledgments}\end{center}
The calculations were carried out on the Xolas prototype SMP cluster.
This work was supported primarily by the MRSEC Program of the National
Science Foundation under award number DMR 94-00334 and also by the
Alfred P. Sloan Foundation (BR-3456).

\end{document}